\documentclass{article}
\usepackage[utf8]{inputenc}
\usepackage{fancybox}
\usepackage{adjustbox}
\usepackage{fancybox}
\usepackage{colortbl}
\usepackage{dcolumn}
\usepackage{amsfonts,amsmath,amssymb}
\usepackage{appendix}
\usepackage{tikz} 
\usetikzlibrary{fit,positioning}
\usepackage[section]{placeins}
\usepackage{lineno}
\usepackage{subcaption}
\usepackage{float}
\usepackage{hyperref}
\usepackage{authblk}
\usepackage{array}
\usepackage{multirow,multicol}
\usepackage{nameref,hyperref}
\usepackage{fancyhdr}

\definecolor{Gray}{gray}{0.9}
\definecolor{Gray1}{gray}{0.7}
\usepackage{adjustbox}
\usepackage{tabularx,colortbl}

\title{The Role of SARS-CoV-2 Testing on Hospitalizations in California}

\author{J. Cricelio Montesinos-L\'opez,
Maria L. Daza–Torres, Yury E. Garc\'ia , Luis A. Barboza, Fabio Sanchez, Alec J. Schmidt, Brad H. Pollock, Miriam Nu\~no}

\begin{document}

\maketitle

%%% First proposal----------------
\begin{abstract}
The rapid spread of the new SARS-CoV-2 virus triggered a global health crisis disproportionately impacting people with pre-existing health conditions and particular demographic and socioeconomic characteristics. One of the main concerns of governments has been to avoid the overwhelm of health systems. For this reason, they have implemented a series of non-pharmaceutical measures to control the spread of the virus, with mass tests being one of the most effective control. To date, public health officials continue to promote some of these measures, mainly due to delays in mass vaccination and the emergence of new virus strains. In this study, we studied the association between COVID-19 positivity rate and hospitalization rates at the county level in California using a mixed linear model.
The analysis was performed in the three waves of confirmed COVID-19 cases registered in the state to September 2021. Our findings suggests that test positivity rate is consistently associated with hospitalization rates at the county level for all waves of study. Demographic factors that seem to be related with higher hospitalization rates changed over time, as the profile of the pandemic impacted different fractions of the population in counties across California.
\end{abstract}

% Keywords
%\keyword{COVID-19; test positivity rate; mixed-effects model} 

\section{Introduction}

The SARS-CoV-2 virus, responsible for the novel coronavirus disease (COVID-19), was identified in late December 2019 in Wuhan, China~\cite{wu2020characteristics}, and spread rapidly, causing a global health crisis. As of October 5, 2021, more than 235 million cases and 4,812,221 deaths have been confirmed  worldwide~\cite{JohnsHopkins}. As the pandemic spread across the globe, governments started to enforce public policies to suppress SARS-CoV-2 transmission, including social distancing, contact tracing, stay-at-home orders, school closings, limits public space utilization, and border closures~\cite{nicola2020health,ebrahim2020covid}. To date, public health officials continue to promote some of these non-pharmaceutical measures, mainly due to delays in mass vaccination and the growing number of new COVID-19 variants~\cite{hadfield2018nextstrain}. Mass surveillance testing, efforts of isolation, quarantine, and contact tracing became essential control measures for curtailing the burden of the COVID-19 pandemic~\cite{bonet2021cost}. The successful epidemic control measures taken by countries such as Korea, Taiwan, Japan, China, New Zealand, and the Czech Republic, which emphasized high testing rates during the initial stages of the pandemic, supported the proposal that mass surveillance testing could help limit viral transmission when properly leveraged~\cite{wang2020response,cao2020post,fang2020large,burki2020mass,louie2021lessons}. However, it remains unknown which testing strategies are the best and whether different approaches show significant and measurable effects on viral spread in general and the rates of severe or deadly cases in particular~\cite{pilecco2021effect}. Although population-scale testing is proven to reduce SARS-CoV-2~\cite{hsiang2020effect}, it appears to become less effective as viral prevalence decreases and is insufficient to eliminate viral transmission on its own~\cite{larremore2020test, pavelka2021impact}.

Public health officials commonly use the test positivity rate to infer the adequacy of population-level testing and the rate of COVID-19 transmission in a population~\cite{Atlantaposrate}. A low test positivity rate indicates low viral prevalence and a testing program with sufficient surveillance capacity. In contrast, a high test positivity rate suggests that the amount of testing is insufficient and that many infected people go unnoticed, especially when test positivity rates are higher than the expected prevalence~\cite{mercer2021testing}. Implementing mass testing may also lead to fewer hospitalizations by reducing new infections by offering interventions for symptomatic and asymptomatic cases discovered early~\cite{neilan2020clinical,wei2021correlation,bonet2021cost}. Hospitalization is also influenced by the demographic structure of the population and health care system factors. In theory, a public health system that is better prepared to identify and support the isolation of cases discovered by surveillance testing and treat those who require medical care should result in 
lower hospitalizations rates.

On January 26, 2020, the first documented case of COVID-19 in California occurred in Orange County~\cite{Wiki}. Since then, the state government has implemented a variety of strategies to contain the spread of the virus~\cite{CAState}. On March 4, 2020, California declared a state of emergency, followed by a mandatory statewide stay-at-home order on March 19, 2020. On June 18, 2021, a statewide mask mandate was ordered due to the rising number of cases and deaths. These mandates were in force until June 15, 2021, when California started reopening the economy~\cite{Governor}, with $70\%$ of eligible with at least one dose of the COVID-19 vaccine and more than $40\%$ of the population fully vaccinated~\cite{CNN}. As of September 22, 2021, California has had three COVID-19 case waves. The first peak occurred in mid-July 2020, reaching an average of 10,000 new cases per day (first wave, May-Sep 2020)~\cite{USAFacts}. During this first wave, most infections were geographically concentrated in the Central Valley, primarily dominated by agriculture, manufacturing services, and retail, meaning few residents could make the transition to working from home~\cite{News3}. In Autumn 2020, COVID-19 cases spiked again, to a peak of 40,000 new cases per day at the end of December (second wave Nov-Jan 2021). During this wave, Los Angeles was one of the main epicenters of the pandemic~\cite{FoxNews,News4}. The third wave associated with the SARS-CoV-2 delta variant started in mid-June 2021 after the lifting of the statewide stay-at-home order. By mid-September, the number of reported daily COVID-19 infections were decreasing, and as of September 20, 2021, California reported the lowest coronavirus state incidences case rate in the U.S.~\cite{News1,News2}

%Using test positivity and hospitalization rates at the county level for the entire COVID-19 pandemic period in California. In order to adjust for differences at the county level that may influence the association between hospitalizations and positivity rate, we consider several factors that will likely confound the relationship of these rates.

In this paper, we aim to provide an exploratory data analysis to verify how demographics and positivity rate correlate with COVID-19 hospital admission in California. The analysis was performed in each of the three waves, using a mixed linear model and data related to hospitalizations for COVID-19, age, race, ethnicity, poverty, and mobility. The paper is organized as follows: Section 2 describes the data and the methods implemented for the analysis, section 3 describes the results, and section 4 highlights the main results and conclusions.

\section{Materials and Methods}
The main goal of this analysis is to describe the effect of surveillance testing on hospitalizations for COVID-19. We performed a comparative analysis using a mixed linear model to study the relationship between hospitalization rates for COVID-19 and positive cases, diagnostic tests, mobility, age, race and ethnic group, poverty, and education across the counties of California. Sixteen of the fifty-eight counties were excluded from this analysis: Calaveras, Colusa, Del Norte, Glenn, Inyo, Lassen, Mariposa, Modoc, Mono, Plumas, San Benito, and Siskiyou, due to low quality of hospital and mobility data; and rural counties like Alpine, Sierra, Sutter, and Trinity because they do not have hospital wards, so patients from those counties would go to neighboring counties for COVID-19 medical care. %(Figure S2)

\begin{figure}[H]
\centering
\includegraphics[scale=0.35]{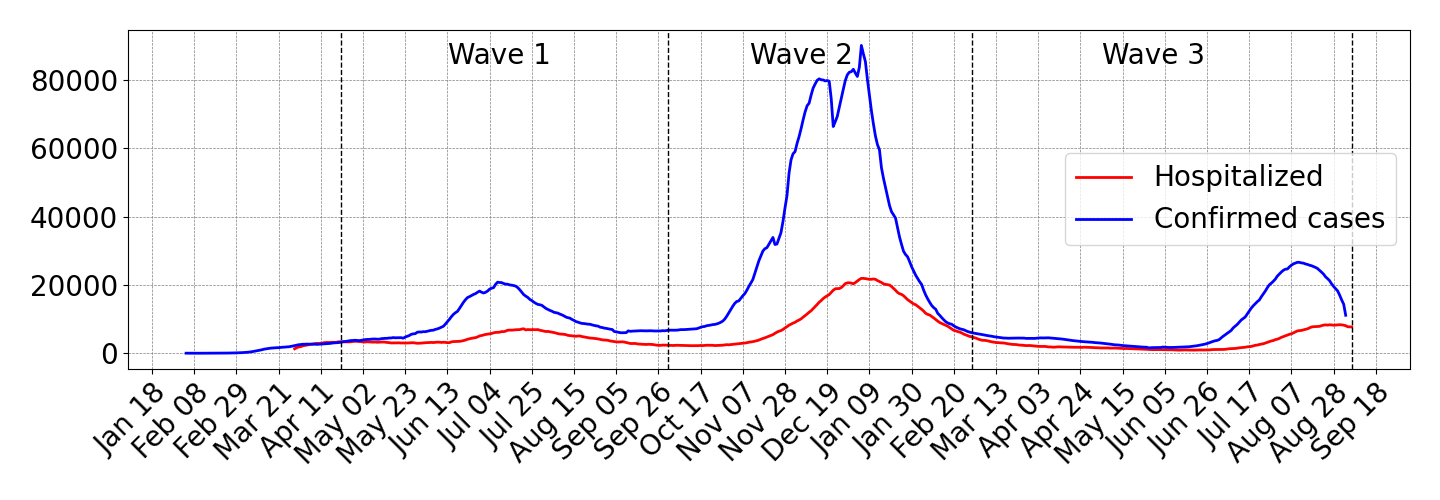}
\caption{Confirmed cases (7-day moving average) and the number of patients hospitalized in an inpatient bed who have laboratory-confirmed COVID-19 in California.}
\label{fig:data_period}
\end{figure}

We analyzed three waves according to the three primary outbreaks reported in California~\cite{CADashboard}. We defined the first wave period from April 21, 2020, to September 30, 2020; the second wave starts on October 1 and ends on February 28, 2021, and the third starts on March 1 and ends on September five, 2021 (Figure~\ref{fig:data_period}). We implemented a change-point analysis to select the different waves intervals.

\subsection{Data Sources}

 Several data sources were used for this study. We used publicly available epidemiological data for COVID-19 daily reported cases and hospitalization admissions at the county level from the official website of the California Department of Public Health (CDPH)~\cite{CDPH}. Data from the American Community Survey (ACS)~\cite{CensusBureau} estimates characteristics at the county level for age and race or ethnic group. We used the Healthy Places Index (HPI) to account for community-level factors contributing to social vulnerability. The HPI is produced by the Public Health Alliance of Southern California, which combines twenty-five community characteristics (e.g., the number of people living below the poverty line, the number of people with lower levels of education, areas with more renters and fewer homeowners, among others) into a single index value to account for the level of poverty, education, and life expectancy in a particular community~\cite{HPIWebsite}. The degree of intra-community mobility was produced from Google's Community Mobility Reports~\cite{Google}. Six Google-specific data streams (grocery and pharmacy, parks, residential, retail and recreation, transit stations, and workplaces) were combined to obtain a single mobility measure for the county using principal component analysis (PCA) (see Supplementary Material for details). All data that changed over time were analyzed weekly to minimize fluctuations observed at the daily level. We considered 7-day averages for daily test positivity rate, intra-community mobility, and hospitalization rate (Figure: S2-S5), given that this is likely to be less volatile.

\subsection{Exposure and Outcome}

The number of tests done and the number of positive cases discovered is not meaningful without extra information. The number of confirmed cases on a given day is related to the actual prevalence, the average duration of disease, and the gross number of tests performed, such that an increase in the number of tests can reveal more existing infections and a change in estimates of the prevalence. Test positivity rate incorporates both the number of tests done and the number of positive cases discovered, frequently used for monitoring the progression of the COVID-19 pandemic~\cite{furuse2021relationship,fenga2021predictive}, and its correlation with hospitalization rates has been shown in previous studies~\cite{kafle2021modeling,al2021positivity} consistent with our use here. We calculated the average positivity rate at the county level by dividing the 7-day average of daily confirmed cases by the 7-day average of daily tests. The hospitalization rate was conceptualized as the average weekly hospital admission rate for laboratory-confirmed COVID-19 per 10,000 county residents. The weekly average positivity and hospitalization rates were log-transformed to capture the effect of detected infections and testing on COVID-19 hospitalizations.

\subsection{Model}

Hospitalization data are made up of repeated measurements. The first, second, and third waves represent 24, 22, and 27 measurements of hospitalization rate, respectively, corresponding to the number of weeks in each wave. The traditional linear regression model is not appropriate for studying data with multiple repeated measures~\cite{10.2307/2529876}. Therefore, we employed a linear mixed-effects model that incorporates repeated observations at the county level.

Let $\textbf{Y}_{j}$ be the $I \times 1$ dependent variable corresponding to the log of the rate of hospital admissions for COVID-19 per 10,000 inhabitants at the county $j$. The subscripts $j=1,2,..., J$ and $i=1,2,..., I$ represent the 42 counties in California and the number of weeks in the wave data collected, respectively. $\textbf{X}_{j}$ is the $I\times p$ fixed-effects design matrix; $\boldsymbol{\beta}$ is the $p\times 1$ fixed-effects vector; $\textbf{Z}_{j}$ correspond to $I\times q$ matrix of random-effects design matrix; $\textbf{u}_{j}$ represents the $q\times1$ vector of random effects and $\boldsymbol{\varepsilon_{j}}$ is the $I\times1$ vector of residuals. $\textbf{u}_{j}$ is independent of $\boldsymbol{\varepsilon_{j}}$. \textbf{G} is the $q\times q$ covariance matrix for the random effects, and $\textbf{R}_{j}$ is the $I\times I$ covariance matrix for the residuals. The model we considered includes a random intercept and a random slope concerning the positivity rate ($q=2)$ since we hypothesize that each county has a different baseline positivity rate and that the effect of the positivity rate on hospitalization differs between counties.

We define the general form of the mixed linear regression model as follows:
\begin{equation}
\begin{split}\textbf{Y}_{j} & =\textbf{X}_{j}\boldsymbol{\beta}+\textbf{Z}_{j}\textbf{u}_{j}+\boldsymbol{\varepsilon_{j}}\\
\textbf{u}_{j} & \sim N\left(0,\textbf{G}\right)\\
\boldsymbol{\boldsymbol{\varepsilon}_{j}} & \sim N\left(0,\textbf{R}_{j}\right).
\end{split}
\label{Eq:Model}
\end{equation}

The term $\textbf{X}_{j}\boldsymbol{\beta}$ corresponds to the fixed effect(s) component (a standard general linear model) and $\textbf{Z}_{j}\textbf{u}_{j}$ to the random effects. The model was fitted using the \textbf{lmer} function in the \textbf{lme4} package for \textbf{R}~\cite{JSSv067i01}.

Since only the hospitalization rate and the positivity rate were log-transformed, we interpret the coefficient ($\beta_r$) for the log positivity rate as the percent increase in the hospitalization rate for every $1\%$ increase in the positivity rate. The estimation for all other coefficients ($\beta_p$'s) requires transformation via $100\times(\exp(\beta_p)-1)$, which gives the percent increase (or decrease) in the hospitalizations rate for every one-unit increase in the independent variable. 

%%%%%%%%%%%%%%%%%%%%%%%%%%%%%%%%%%%%%%%%%%
\section{Results}

The coefficient estimates and the $95\%$ confidence intervals (CI) for the linear mixed model are presented in Table~\ref{Tab:Results1}. The $\beta$ value represents the effect that each variable has on the hospitalizations rate. Variables with a p-value $<0.05$ were considered statistically significant. Results show that significant variables changed over time, but the positivity rate consistently remained significant across all three waves with a coefficient $\beta_r$ close to one. Regarding hospitalization rates for different racial and ethnic groups, counties with a higher population percentage of non-White race or ethnic groups had higher hospitalization rates in the first and second waves~\ref{Tab:Results1}. In the first wave of infections, counties saw an average $7.4\%$ increase in hospitalization rate for every $1\%$ of the population identified as Hispanic or Latino, and a $16.6\%$ increase in hospitalization rate for every $1\%$ of the population that identifies as African American. In the second wave, counties with high proportions of Hispanic or Latino and African American populations were not significantly different, but a $3.4\%$ increase in hospitalization rates was associated with every $1\%$ of the population that identifies as Asian.

HPI was significant and positive in the first wave, meaning that counties with more significant economic, social, and healthcare resources reported increased hospitalization rates compared to counties with fewer resources. Higher intra-community mobility was associated with higher hospitalization rates; however, in the second wave, we found that higher mobility was negatively associated with hospitalization rates. 

\begin{table}[H]
\caption{Association between hospitalization rates and independent variables at the county level.}
\resizebox{\columnwidth}{!}{%
\begin{tabular}{|l|c|c|c|c|c|c|c|}
\rowcolor{Gray}\hline
 & \multicolumn{2}{c|}{{\bf 1st wave }}   & \multicolumn{2}{c|}{{\bf 2nd wave }}  &  \multicolumn{2}{c|}{{\bf 3rd wave}}\\
 \rowcolor{Gray}\hline
{\bf Variable} & {\bf Estimate* (95\% CI)} & {\bf p-value} & {\bf Estimate* (95\% CI)} & {\bf p-value} & {\bf Estimate* (95\% CI)} & {\bf p-value} \\ 
\hline
 Positivity rate & 0.9 (0.6, 1.4) & $<$0.001 & 0.9 (0.8, 0.9) & $<$0.001 & 1.1 (0.9, 1.2) & $<$0.001 \\
 Pop over 65 & -13.2 (-30.6, 8.5)  & 0.246 & 3.1 (-3.8, 10.6) & 0.416 & -7.7 (-15.1, -0.1) & 0.060\\
 Asian &  0.8 (-7.2, 9.4) & 0.861 & 3.4 (0.7, 6.1) & 0.020 & -0.5 (-3.9, 2.9) & 0.763\\ 
 Hispanic/Latino & 7.4 (2.1, 12.9) & 0.010 & 0.9 (-0.6, 2.5) &
 0.259 & -0.6 (-2.3, 1.3) & 0.535 \\ 
 African American & 16.6 (0.1, 35.9) & 0.070 & 0.1 (-4.7, 5.1) & 0.982 & 1.8 (-4.3, 8.2) & 0.582 \\ 
 HPI & 4.9 (2.4, 7.4) & $<$0.001 & -0.4 (-1.2, 0.3) & 0.266 & 0.2 (-0.7, 1.2) & 0.686 \\
 Mobility & 4.9 (1.9, 7.9) & $<$0.001 & -2.8 (-3.4, -2.1) & $<$0.001 & 0.1 (-1.7, 1.8) & 0.950 \\
 \hline
\end{tabular}
}
\footnotesize{*A $1\%$ increase in the positivity rate coefficient consistently corresponds to around a $1\%$ percent increase in the hospitalization rate. Interpretation for the other independent variables is a one-unit increment corresponding to a percent change, namely positive or negative, depending on the coefficient sign.}
\label{Tab:Results1}
\end{table}

Table~\ref{Tab:Results2} displays the coefficient value related to the log positivity rate for each county in the three waves. These values are equal to ($\beta_r + u_{r_j}$), where $\beta_r$ correspond to the general coefficient for the log positivity rate (Table~\ref{Tab:Results1}) and $u_{r_j}$ is the random coefficient for the j-th county, $j=1,2,..., J$. Counties with higher coefficient values had stronger associations between test positivity rate and hospitalization rate.

\begin{table}[H]
\centering
\caption{Estimates of the association between test positivity rate and hospitalization rate for each wave of study.}
\label{Tab:Results2} 
\resizebox{\columnwidth}{!}{%
\begin{tabular}{|l|l|l|l|l|l|l|l|}
\rowcolor{Gray}\hline
County	&	Wave 1	&	Wave 2	&	Wave 3	&	County	&	Wave 1	&	Wave 2	&	Wave 3\\
\hline
Yolo	    &	2.99 	&	0.70 	&	1.10 	&	
Santa Clara	&	0.95      	&	1.08 	&	0.82 \\
Madera	    &	1.64 	&	0.77 &	1.05 	&	
Stanislaus	&	0.92 	&	0.80 	&	0.78 \\
El Dorado	&	1.52 	&	1.34 	&	1.53 	&	
Nevada	    &	0.92 	&	0.64 	&	0.94 \\
Imperial	&	1.34 	&	1.02 	&	0.92 	&	
Tuolumne	&	0.91 	&	0.55 	&	0.80 \\
Los Angeles	&	1.23 	&	0.95 	&	0.78 	&	
Butte	    &	0.90 	&	0.82 	&	1.10 \\
Orange	    &	1.22 	&	1.02 	&	0.94 	&
Merced	    &	0.90 	&	0.86 	&	1.30 \\
Alameda	    &	1.18 	&	0.82 	&	0.98 	&
San Francisco&	0.82 	&	1.06 	&	0.87 \\
Tulare	    &	1.15 	&	0.86 	&	1.00 	&	
Marin	    &	0.79 	&	0.79 	&	0.98 \\
Fresno	    &	1.14 	&	0.88 	&	0.87 	&	
Sacramento	&	0.75 	&	0.85 	&	1.16 \\
Contra Costa	&	1.09 	&	0.99 	&	1.21 	&	
Napa	        &	0.73 	&	0.95 	&	1.04 \\
San Bernardino	&	1.06 	&	1.01 	&	1.02 	&	
Lake         	&	0.71 	&	1.06 	&	2.62 \\
Kern	        &	1.06 	&	0.84 	&	1.05 	&	
Amador      	&	0.71 	&	0.90 	&	1.01 \\
San Mateo	    &	1.04 	&	1.03 	&	0.91 	&
Shasta      	&	0.70 	&	1.02 	&	0.91 \\
Ventura	        &	1.04 	&	1.05 	&	0.91 	&	
Tehama	        &	0.69 	&	0.76 	&	1.70 \\
Santa Barbara	&	1.04 	&	0.93 	&	0.97 	&	
Sonoma	        &	0.67 	&	0.97 	&	1.03 \\
San Joaquin	    &	1.01 	&	1.01 	&	1.06 	&	
Santa Cruz	    &	0.66 	&	0.79 	&	1.12 \\
Kings	        &	1.00 	&	0.77 	&	0.89 	&	
San Luis Obispo	&	0.55 	&	0.95 	&	1.15 \\
San Diego	    &	0.99 	&	0.92 	&	0.97 	&	
Yuba	        &	0.52 	&	0.95 	&	1.02 \\
Solano	        &	0.99 	&	0.90 	&	1.06 	&	
Placer       	&	0.29 	        &	0.89 	&	0.77 \\
Riverside	    &	0.97 	&	1.02 	&	1.09 	&	
Mendocino    	&	0.18 	&	1.10 	&	1.17 \\
Monterey	    &	0.95 	&	0.81 	&	1.05 	&	
Humboldt	    &	0.12 	&	0.38 	&	1.14 \\
\hline
\end{tabular}}
\end{table}

%%%%%%%%%%%%%%%%%%%%%%%%%%%%%%%%%%%%%%%%%%%%%%%%%%%

\section{Discussion}

A mixed linear model was used between the COVID-19 hospitalization rate and factors such as age, ethnicity, race, poverty index, and intra-community mobility. Our primary interest was studying the impact of testing rates on county-level hospitalization rates, as county health departments were usually responsible for public testing administration. We found that the test positivity rate was consistently significant and positively associated with the hospitalization rate during all three waves of COVID-19. Hospitalization rate increased at an almost 1:1 basis with a positivity rate. While other possible predictors of hospitalization rate, including the density of different race or ethnic groups, social vulnerability, and intra-community mobility, had pronounced effects at differing times during the pandemic, none were consistent predictors of hospitalization rate for all three waves of infection.

The actual local prevalence and the number of tests administered both affect the positivity rate value. Generally, the higher the true prevalence, the higher the positivity rate will be; as more tests are deployed, the positivity rate will converge with the true prevalence. The nature of diagnostic testing on a first-come-first-serve basis frequently leads to positivity rates more than the actual prevalence if testing rates are insufficient to sample the mild or asymptomatic cases. In other words, if the number of tests is a limiting factor, and they are used primarily to confirm likely cases more frequently than a random surveillance sampling of the population, positivity rates will be biased upwards compared to the actual prevalence. This assumes that those who suspect they have the disease or suspect exposure are more likely to seek a test than those who have no such suspicion. Thus, high test positivity rates are likely a mix of biased sampling and high prevalence, but clarifying which is dominant during a specific time frame requires high-quality auxiliary data that may not exist. Our results suggest that actions that reduce the test positivity rate are likely to reduce the hospitalization rate by a similar magnitude. Simply increasing the number of tests will only significantly reduce the positivity rate if sampling bias is the dominant reason for a high positivity rate. Determining the effect on hospitalization rate of reducing test positivity rate in bias-dominant versus prevalence-dominant systems is beyond the scope of this paper, but remains an important question.

The response following detection is essential. Theoretically, early detection of a new case, symptomatic or asymptomatic, and rapid isolation will prevent further potential hospitalizations. Extrapolating from our results, we expect that the counties that more regularly tested a more significant proportion of their population- from asymptomatic surveillance or robust testing requirements for essential workers- experienced lower hospitalization rates than the counterfactual scenario. However, care must be taken extending this reasoning too far: large-scale population testing can theoretically lead to reduced hospitalizations, but the effect will always be indirect. The resources and infrastructure must support proper mass testing and preparation to respond to the information garnered from the testing program, which no two counties will have done identically, hence why each county reported here maintained intercepts that varied from each other over time.   

A low positivity rate due to a high amount of testing does not always imply adequate pandemic control. Not only does the gap between testing rates among suspected cases, known exposures, and the unexposed or asymptomatic matter, but testing rates among different demographic groups demonstrably effect the value of testing data. Suppose the mass testing systemically excludes people with a high-risk profile (as could quickly occur where healthcare accessibility is low). In that case, many infections could remain undiscovered for long periods, leading to a growth in the hospitalization rate despite low positivity rates. The pandemic has not affected everyone equally. Disparities in coronavirus disease outcomes by racial and ethnicity as well as socioeconomic status have been reported since the beginning of the pandemic~\cite{chowkwanyun2020racial}. Our findings highlight that areas with larger relative populations of Hispanic or Latinos and African Americans were significantly correlated with higher hospitalization rates in the first wave and with Asians in the second wave, consistent with previous studies~\cite{escobar2021racial,ogedegbe2020assessment,podewils2020disproportionate}. The underlying causes of health disparities in Latinos, African Americans, and poor communities are related to social and structural determinants of health~\cite{hooper2020covid}. Implementing social distancing, especially at the beginning of the pandemic, may have been challenging because these communities, on average, live in more crowded conditions and work more frequently in essential public-facing occupations. In addition, their access to health services is systemically limited, so that populations have a disproportionate burden of underlying comorbidities and lack the possibility of accessing adequate and timely treatment when affected by the SARS-CoV-2 virus~\cite{bibbins2020time}, and possibly confounding the relationship between test positivity rates and hospitalization rates, as discussed above.

The Healthy Places Index is correlated positively with the hospitalization rate in the first wave, which implies that counties with higher socioeconomic status had a higher probability of reporting hospitalizations. One of the reasons may be the capacity and better availability of
hospital facilities attributed to economic resources. Mobility was another significant variable that positively and negatively correlated with the hospitalization rate in the first and second waves. A similar result was reported in~\cite{bhowmik2021comprehensive} for COVID-19 transmission and mortality rates. Early in the pandemic, mobility patterns were drastically affected by containment measures implemented to slow the spread of the disease. Our results show a linear correlation between mobility and the rate of hospitalization in the first wave, in agreement with previous reports~\cite{gottumukkala2021exploring}, which implies that an increase in the circulation of people could cause an increase in infections and, consequently, in hospitalizations. However, it is not clear how mobility affected the growth rates of the COVID-19 infection once the lockdowns were lifted because other interventions became more widely available and easier to adhere to, such as wearing face masks and social distancing, patterns of both mobility and growth of infections became non-linear~\cite{gottumukkala2021exploring}. One interpretation could be that areas with lower infection rates allowed for greater freedom in summer activities, negatively correlating positivity and hospitalization rates. Care must be taken in attributing causation to relationships between these covariates and hospitalization rates without further study.

This study has some limitations that are important to consider. First, it is focused on county-level analysis and is intended to investigate population-level risk; conclusions at the individual level are not appropriate and should not be applied. Second, as discussed earlier, we did not attempt to address whether a given data point on test positivity was produced during a bias-dominant or prevalence-dominant period. Thus an unknown proportion of the relationship between positivity rate and hospitalization rate is likely due to natural increases in the prevalence. Third, the hospitalization rate is also dependent on available hospital beds, which we did not consider as a factor given the limited availability and reliability of such data at the county level. Thus, some instances where hospitalization would have been an outcome for a patient except for bed availability were not accounted for, which could have led to point underestimates of our primary outcome measurements.

Knowing the factors that affect the spread of the virus and hospitalizations allows local decision-makers to help identify areas at higher risk for severe COVID-19 and guide resource allocation and implementation of prevention and mitigation strategies. These findings highlight how the most significant factors impacting hospitalizations have changed with the pandemic's evolution. The positivity rate is the only factor to prevail over time as a significant and directly correlated with hospitalization rate.

\bibliographystyle{unsrt}
\bibliography{biblio.bib} 
\newpage
%% optional
%\supplementary{}
\section*{Supplementary material} 

\subsection*{Maps}
This maps displays the coefficient value related to the log positivity rate for each county in the three waves.
\begin{figure}[H]
\centering
\includegraphics[scale=0.55]{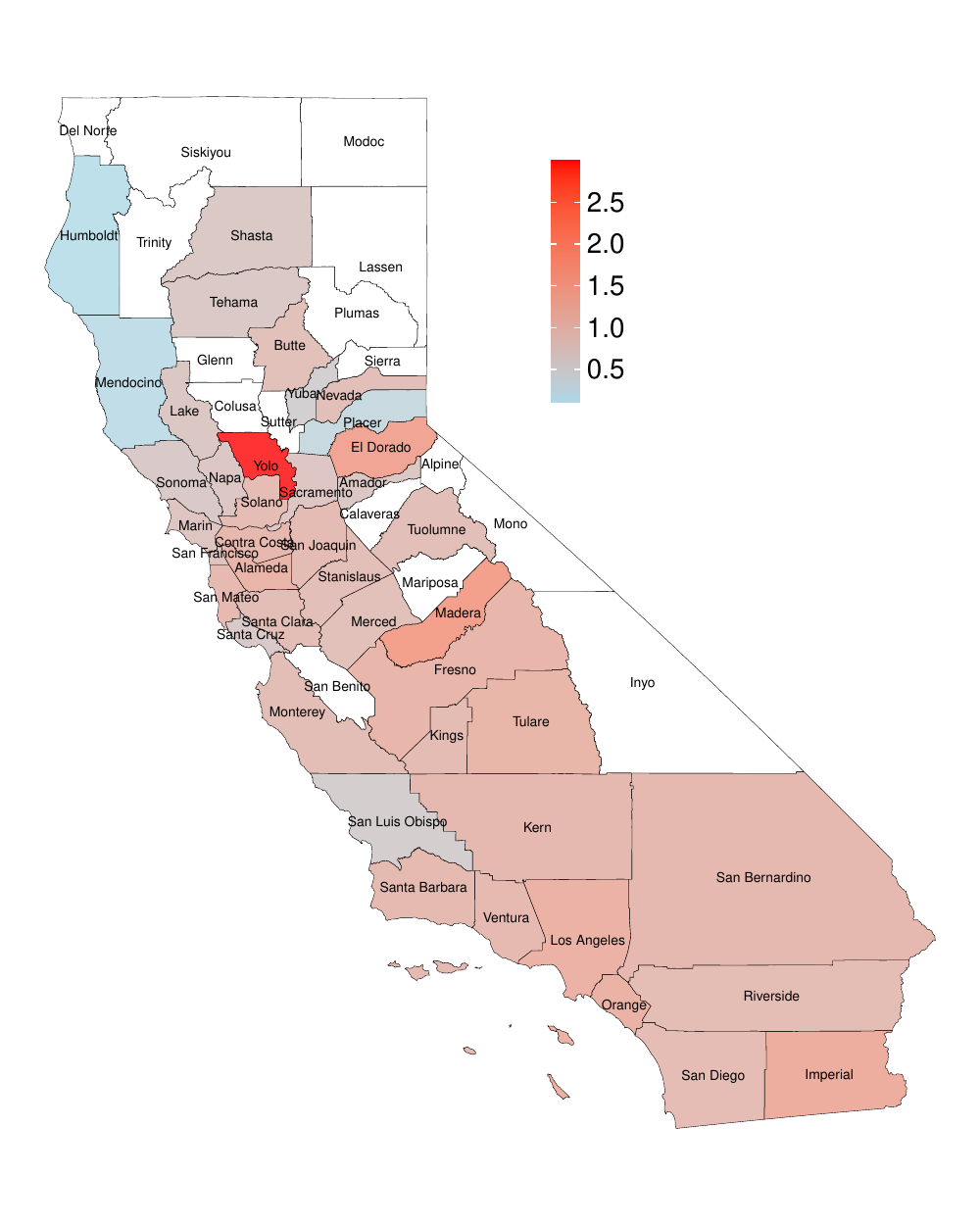}
\includegraphics[scale=0.55]{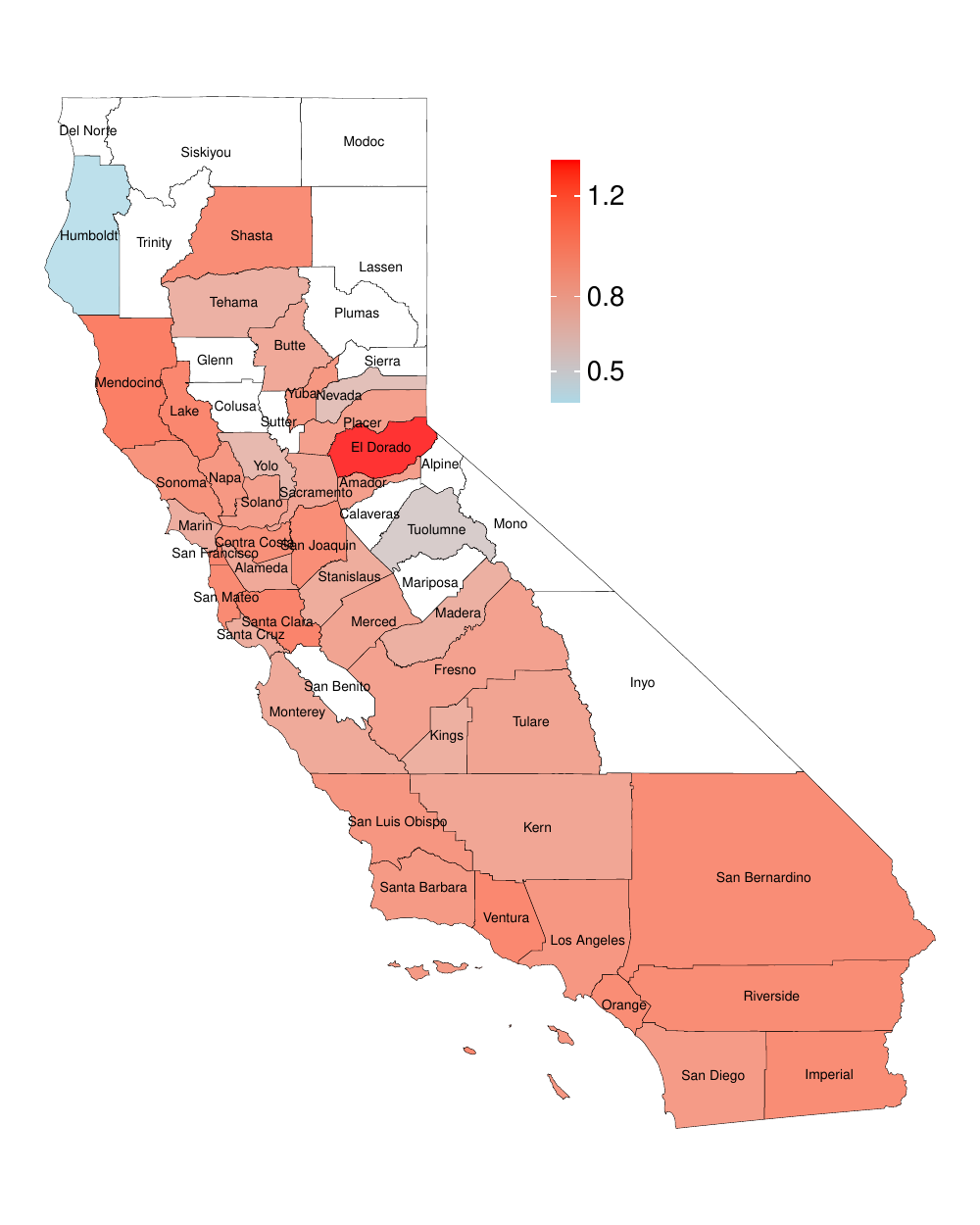}
\includegraphics[scale=0.55]{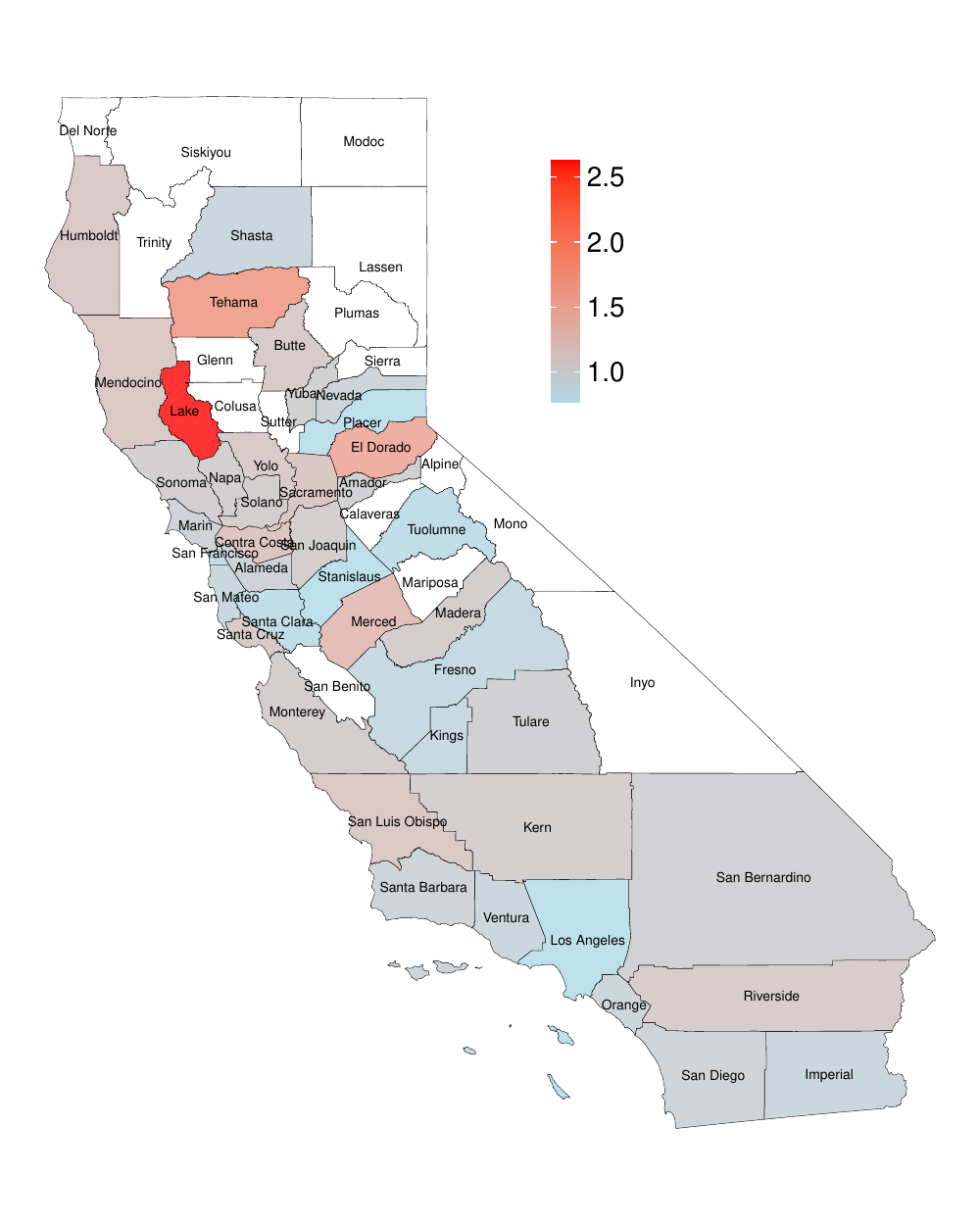}
\caption*{\textbf{Figure S1.} Values in the county-level maps represent the effect of the positivity rate on hospitalizations at each county for each wave. Top Left: Wave 1. Top Right: Wave 2. Bottom: Wave 3.}
\label{fig:Maps}
\end{figure}

\subsection*{Hospitalizations, testings, and positivity rate.}
\begin{figure}[H]
\centering
\includegraphics[scale=0.48]{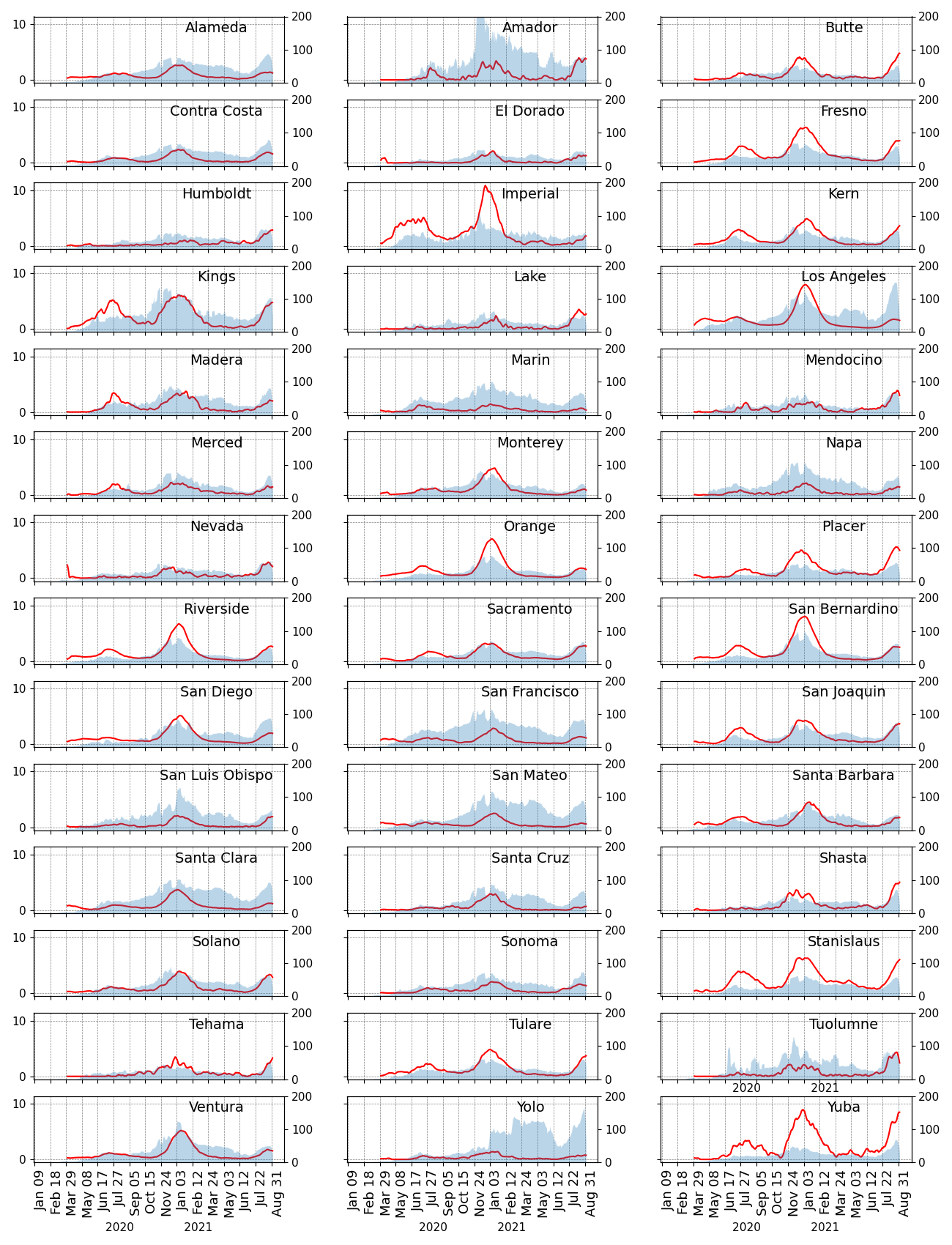}
\caption*{\textbf{Figure S2}. This figure displays the number of patients hospitalized in an inpatient bed who have laboratory-confirmed COVID of the California counties \cite{CDPHHosp} considered in the study and testing per 10K population from January 5, 2020, to September 6, 2021. 
\label{fig:hosp_testing}}
\end{figure}

\begin{figure}[H]
\centering
\includegraphics[scale=0.5]{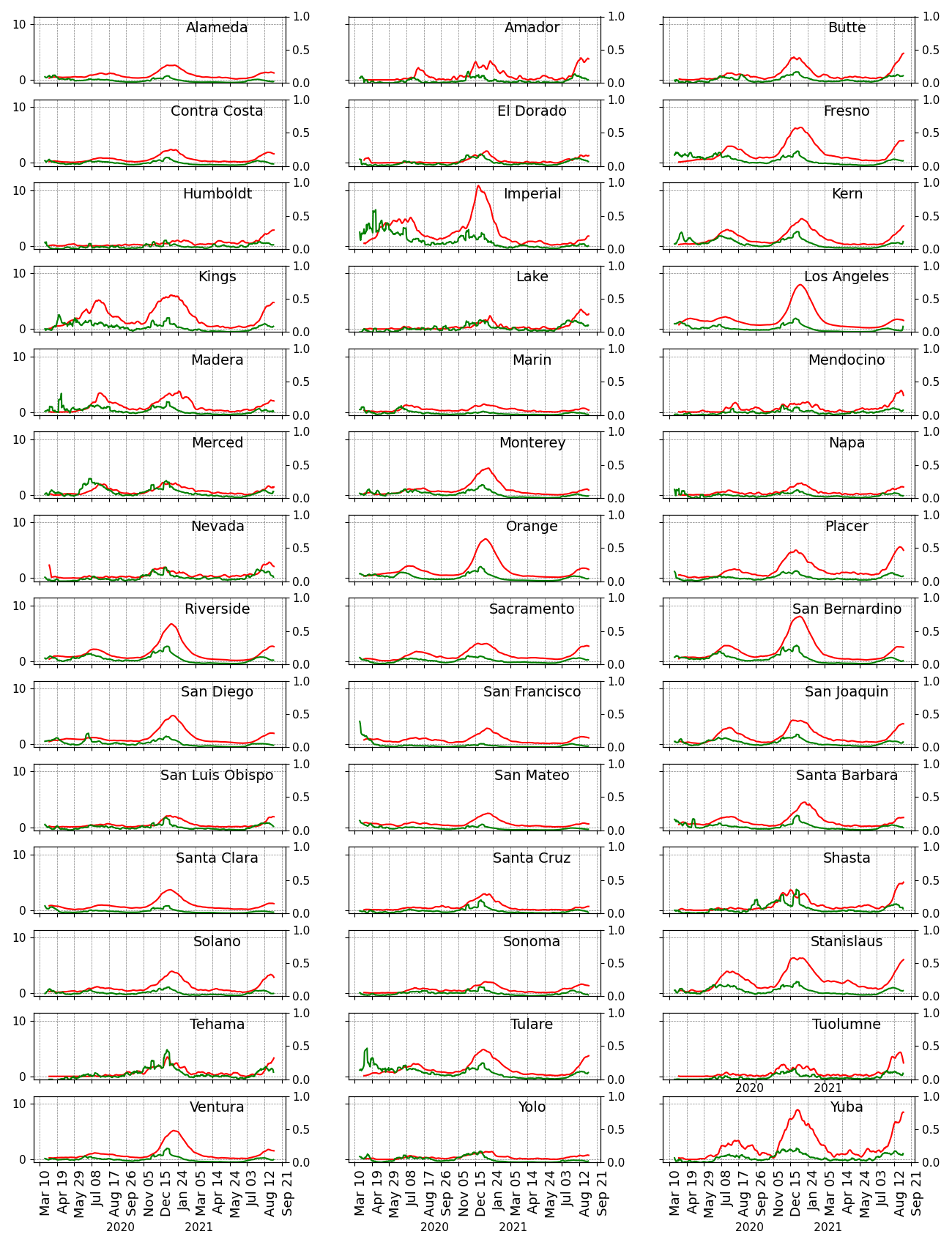}
\caption*{\textbf{Figure S3}. Number of patients hospitalized in an inpatient bed who have laboratory-confirmed COVID (red line) and positivity rate (7-day moving average, green line) from March 28, 2020 to September 6, 2021. 
\label{fig:hosp_post_rate}}
\end{figure}

\subsection*{Variable description}

We describe the co-variable used in the analysis, mobility,  age, race, ethnicity, poverty, and education in the California counties.  

\subsubsection*{Demographic variables}
Demographic characteristics such as age, ethnicity, and race by county was obtained from the  United State Census Bureau~\cite{CensusBureau}.

\begin{table}[H]
\centering
\caption*{\textbf{Table S1}. Demographic variables considered in the mixed linear model.}
\resizebox{\columnwidth}{!}{%
\begin{tabular}{lll}
\hline
\rowcolor{Gray} {\bf Variable}        &  {\bf Description}  \\
%\rowcolor{Gray}
Over 65& Percentage of population ages 65 and above\\
Black or African & Percentage of Black or African American population\\
American Indian /and Alaska Native & Percentage of American Indian Alaska Native population \\
Asian& Percentage of Asian population\\
Hispanic or Latino& Percentage of Hispanic or Latino population\\
HPI & Healthy place index\\
\hline
\end{tabular}}
%\label{tab:var_descrip}
\end{table}

\begin{figure}[H]
\centering
\includegraphics[scale=0.55]{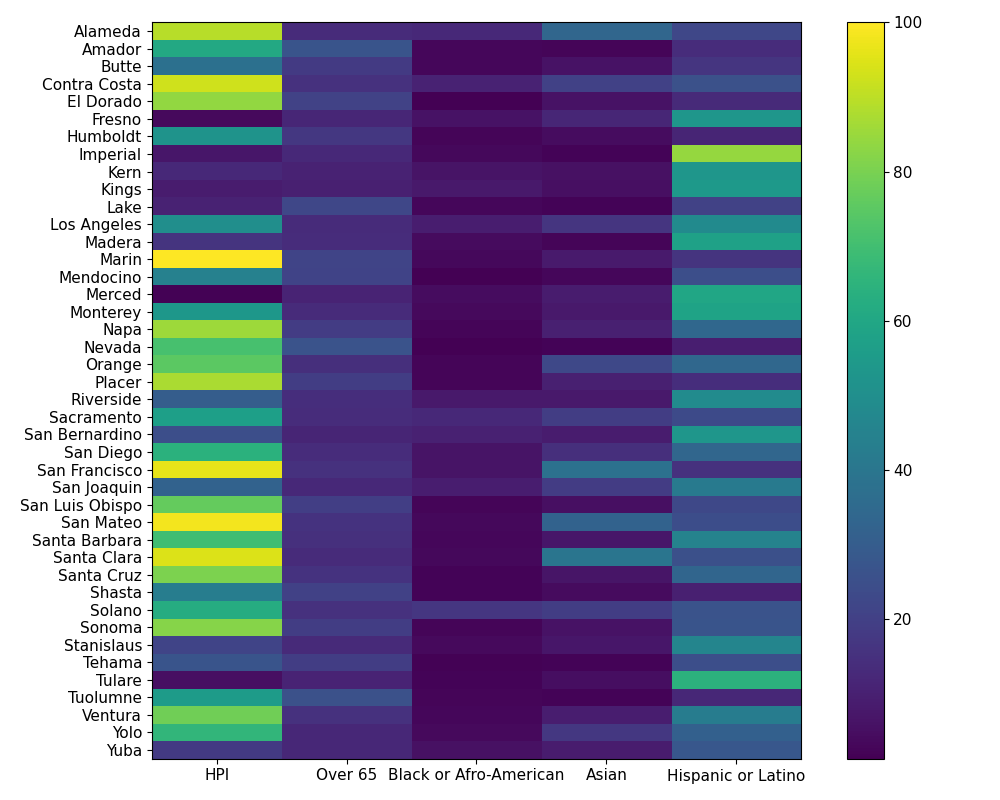}
\caption*{\textbf{Figure S4}. Demographic variables to county level are taken from the United States Census Bureau~\cite{CensusBureau}. All the values are in percentage on the population except the HPI~\cite{HPIWebsite}, which is a number between 0 to 100.\label{fig:demo}}
\end{figure}

\subsubsection*{Correlation and multicollinearity}

To choose the variables used in the model, we checked for the multicollinearity issue across the independent variables as some county-level features may be highly correlated. We plot a correlation matrix to highlight the Pearson correlation coefficient between each independent variable in the dataset. Figure S5 highlights the weaker correlation across most demographics variables while finding high correlations of the comorbidities between them and most of the demographic variables. 

\begin{figure}[H]
\centering
\includegraphics[scale=0.55]{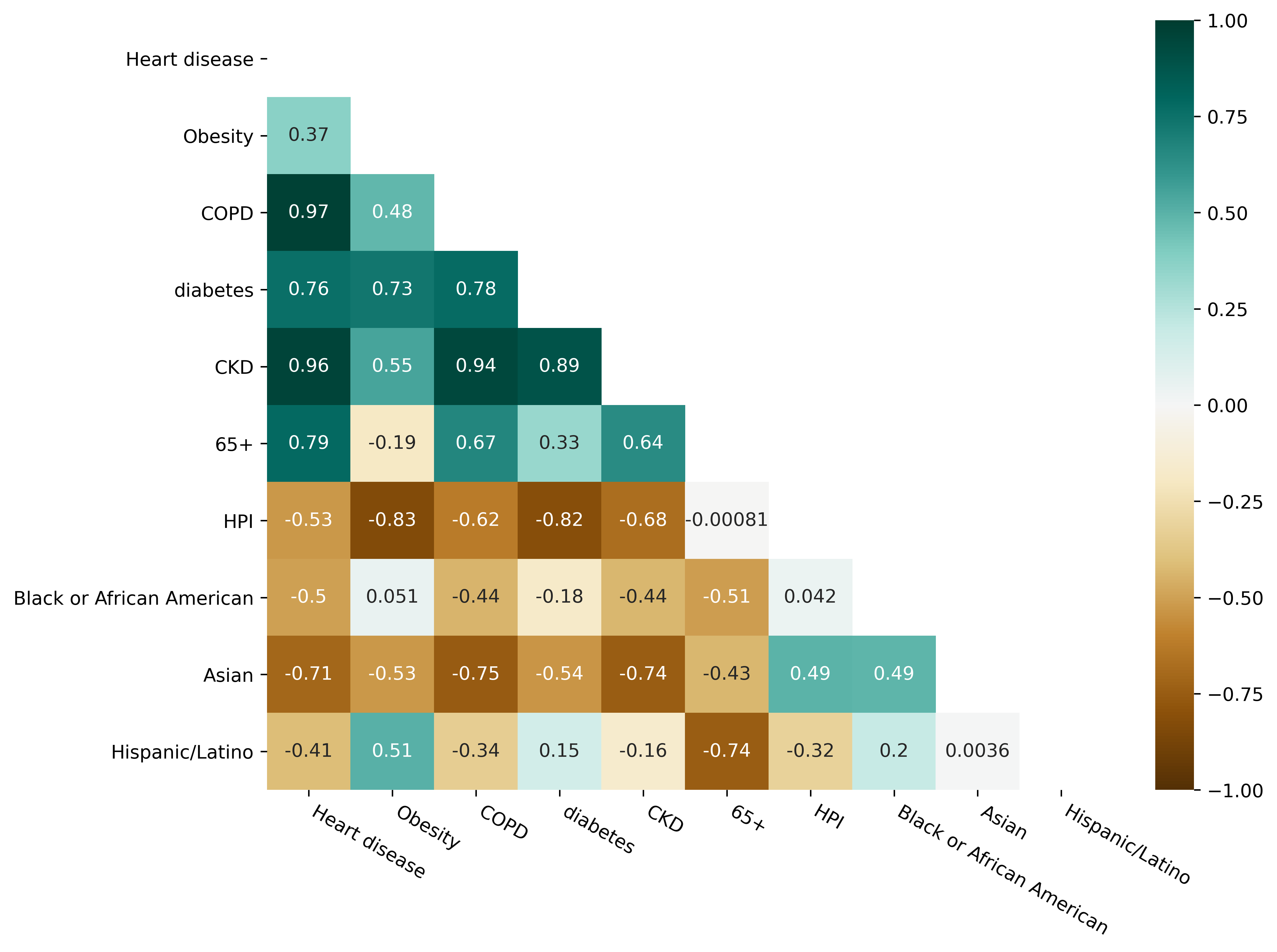}
\caption*{\textbf{Figure S5}. Correlation matrix of the demographics variables and comorbidities.}
%\label{fig:corr_map}
\end{figure}

To measures the severity of multicollinearity in our regression analysis, we estimate the variance inflation factor (VIF) for the previous variables. VIF indicates the increase in the variance of a regression coefficient for each variable as a result of correlated predictors. A VIF value of 1 for a variable implies this variable is not correlated to the remaining ones. When VIF is higher than 10 there is significant multicollinearity that needs to be corrected. In Tab. S2, we observe very large values of the VIF. When removing variable by variable we are left with only the demographic variables given in the Tab. S1. Therefore we can safely conclude that our analysis was unaffected by the multicollinearity issue. For this reason, disease prevalence variables were not included in the analysis.

\begin{table}[H]
\centering
\caption* {\textbf{Table S2}. Variance inflation factor.}
\begin{tabular}{ll} \rowcolor{Gray} \hline
Variable &      Variance inflation factor \\ 
Heart disease &  3846.96 \\
Obesity &   235.79 \\
COPD &  1648.60 \\
diabetes &  1430.87 \\
CKD &  3342.10 \\
65+ &   299.79 \\
HPI &    11.71 \\
Black or African American & 6.50 \\
Asian &    11.19 \\
Hispanic or Latino &    73.89 \\ \hline
\end{tabular}
%\label{tab:VIF}
\end{table}

\subsubsection*{Mobility trends}

The existence of social contacts could be proxied by mobility data \cite{nouvellet2021reduction,badr2020association,bergman2020mobility,gatalo2021associations}, with frameworks such as Google’s Community Mobility Report (CMR) \cite{Google}, COVID‑19 - Mobility Trends Reports - Apple \cite{Apple} and Safegraph \cite{SafeGraph} being able to measure mobility, as it measures citizens’ mobility according to different types.

Our analysis is based on 40 counties of California for which both hospitalization and Google mobility data were available. Google mobility data included six data-streams: “grocery and pharmacy,” “parks,” “residential,” “retail and recreation,” “transit stations,” and “workplaces.” We combined all Google-specific data streams to obtain a google county mobility measure. We used an unsupervised machine learning method known as principal component analysis (PCA) to construct the google mobility index using the six mobility metrics.The first principal component explained more than  50\% of the variability in the data by each county, indicating a good dimension reduction (Table S3).

A regression analysis was used to estimate the lag length. The results show that mobility is  correlated with COVID-19 hospitalizations in most counties with lags of 3-4 weeks.

\begin{table}[H]
\centering
\caption*{\textbf{Table S3}. The variance explained by the first principal component for the Google’s Community Mobility Report (CMR). } 
\resizebox{\columnwidth}{!}{%
\begin{tabular}{|p{3cm}|p{2cm}||p{3cm}|p{2cm}||p{3cm}|p{2cm}|} 
\rowcolor{Gray}\hline
{\bf County}   &  {\bf Explained variance} & {\bf County}   &  {\bf Explained variance} & {\bf County}   &  {\bf Explained variance}\\
Alameda         & 0.62  &Mendocino          & 0.60  & San Mateo       & 0.65 \\
Amador          & 0.43  & Merced           & 0.68 & Santa Barbara   & 0.67\\
Butte           & 0.56  & Monterey         & 0.66 & Santa Clara    & 0.60\\ 
Contra Costa    & 0.63  & Napa             & 0.56 & Santa Cruz     & 0.60 \\ 
El Dorado       & 0.50  & Nevada           & 0.58 & Shasta         & 0.51 \\ 
Fresno          & 0.67  & Orange           & 0.68 & Solano         & 0.60\\ 
Humboldt        & 0.60  & Placer           & 0.50 & Sonoma         & 0.66\\   
Imperial        & 0.67  & Riverside        & 0.65 & Stanislaus     & 0.62\\ 
 Kern            & 0.60 & Sacramento      & 0.66& Tehama         & 0.4\\ 
 Kings           & 0.61 & San Bernardino  & 0.62 & Tulare         & 0.57\\   
Lake            & 0.65  & San Diego       & 0.71 & Tuolumne    & 0.47  \\
Los Angeles     & 0.73  & San Francisco    & 0.73 & Ventura        & 0.56  \\ 
Madera          &  0.51 & San Joaquin      & 0.68 & Yolo           & 0.66\\ 
Marin           & 0.62  & San Luis Obispo  & 0.65 & Yuba           & 0.54\\ 
\hline \end{tabular} }
%\label{tab:variance} 
\end{table} 

\end{document}